\documentstyle[12pt]{article}

\def\be{\begin{equation}}
\def\ee{\end{equation}}
\def\a{\alpha}

\def\s{\sigma}

\def\d{\partial} 
\def\l{\lambda} 
\def\det{\mbox{det}}
\def\ln{\mbox{ln}}
\def\exp{\mbox{exp}}
\def\cos{\mbox{cos}}
\def\sin{\mbox{sin}}
\def\ak{a^{+}}
\def\ck{c^{+}}
\def\ra{\rangle}
\def\la{\langle}
\def\Nt{\tilde{N}}
\def\D{\Delta}
\def\n{\tilde{n}}
\def\r{\rho}

\def\xp{x^{\prime}}
\def\yp{y^{\prime}}

\begin{document}

\begin{center}
{\bf On the universal relations for the formfactors in 1D quantum liquids.} 
\end{center}
\vspace{0.2in}
\begin{center}
{\large A.A.Ovchinnikov}
\end{center}

\begin{center}
{\it Institute for Nuclear Research, RAS, Moscow}
\end{center}

\vspace{0.2in}

\begin{abstract}

For various one-dimensional quantum liquids in the framework of the Luttinger model 
(bosonization) we establish the relations between the coefficients before the 
power-law asymptotics of the correlators (prefactors) and the formfactors of the 
corresponding local operators. The derivation of these relations in the framework 
of the bosonization procedure allows to substantiate the prediction for the formfactors 
corresponding to the low-lying particle-hole excitations. We present an explicit 
expressions for the particle-hole formfactors for various one-dimensional models. 
We also obtain  the formulas for the summation over the particle-hole states 
corresponding to the power-law asymptotics of the correlators.

\end{abstract}

\vspace{0.2in}

{\bf 1. Introduction}

\vspace{0.2in}

Calculation of correlation functions in the 1D quantum liquids or 
spin systems remains important problem both from theoretical and 
experimental points of view. Although the predictions of the 
critical exponents corresponding to the power-law decay at large 
distances obtained with the help of the mapping to the Luttinger model
(bosonization) \cite{LP},\cite{H} or conformal field theory \cite{C}, 
\cite{MZ} are available for a long time, the calculation of the constants 
before the asymptotics (prefactors) remains an open problem. 
Recently some progress was achieved in relating of the prefactors 
to the certain formfactors of local operators by means of direct 
calculations in the integrable model \cite{S1} and by means of the 
conformal field theory \cite{G}. Moreover, for the XXZ - quantum spin 
model the behaviour of the formfactors for the low-lying particle-hole 
excitations was found \cite{S2}, which agrees with the predictions of 
the recent papers \cite{G}.

In Ref.\cite{G} the arguments on the particle-hole formfactors are 
based on the proportionality of the correlator to the certain correlator 
in the Luttinger liquid theory. Although these arguments could in 
principle lead to conclusions made in these papers 
(proportionality of the two correlators means the correspondence 
of the matrix elements and the momenta for the intermediate states) 
the detailed explanation of the results for the particle-hole 
formfactors is absent. Therefore their results on the particle-hole 
formfactors are not grounded enough. It is the main goal of the present 
paper to present the detailed derivation of the results for the 
particle-hole formfactors in the framework of the bosonization approach. 
Next, in Ref.\cite{G} the scaling relations for the lowest formfactors 
are derived using the conformal field theory. However, clearly, 
it would be successive to obtain the results for both type of the 
formfactors within the same method. 
Thus it is desirable to derive the scaling relations for the lowest 
formfactors entirely using the bosonization technique without use of the 
conformal field theory. 
Thus the second goal of the 
present paper is to derive the scaling relations for the lowest 
formfactors in the framework of the bosonization approach. 
To achieve these goals we introduce the extended bosonization concept 
(see the end of Section 2). 
Next we present an explicit expressions fo the 
particle-hole formfactors for different operators for various 
one-dimensional models. This goal can also be achieved only in the 
framework of the bosonization technique. 
We also point out the summation formulas which can be used 
to calculate the sum over the intermediate states for the correlators 
in the framework of the approach of ref.\cite{S2} to the correlators in the 
integrable (XXZ- spin chain) model.  
Finally, we present the results for the particle-hole formfactors 
for the excitations corresponding to the left Fermi-point. 
The short version of the present paper was published in Ref.\cite{O1}. 
 
In Section 2 we fix the notations and briefly review the theory of 
Luttinger liquid and the bosonization procedure. In Section 3 we explain 
how to derive the scaling relations for the lowest formfactors using the 
bosonization approach. Finally in Section 4 we derive the particle-hole 
formfactors for the low-lying states and present an explicit expressions 
for the formfactors of various operators for different models. 
We also present the summation formulas relevant to the calculation of 
the correlators.

\vspace{0.2in}

{\bf 2. Bosonization.}

\vspace{0.2in}

Consider the effective low -energy Hamiltonian which build up from 
the fermionic operators ($a_k,~c_k,~k=2\pi n/L,~n\in Z$, L- is the length 
of the chain) 
\[
a^{+}(x)=\frac{1}{\sqrt{L}}\sum_{k}e^{-ikx}a_{k}^{+}, ~~~~~
a^{+}_k=\frac{1}{\sqrt{L}}\int_{0}^{L}dx e^{ikx}a^{+}(x), 
\]
corresponding to the excitations around the right and the 
left Fermi- points and consists of the kinetic energy term and the 
interaction term $H=T+V$ with the coupling constant $\l$:    
\be
H=\sum_{k}k(a_k^{+}a_k-c_k^{+}c_k)+2\pi\l/L\sum_{k,k',q}\ak_ka_{k+q}\ck_{k'}c_{k'-q}.      
\label{ham}
\ee
Defining the operators \cite{ML}
\[
\r_1(p)=\sum_k\ak_{k+p}a_k,~~~~\r_2(p)=\sum_k\ck_{k+p}c_k,
\]
where $|k|,|k+p|<\Lambda$, where $\Lambda$ is some cut-off energy, 
which for the states with the filled Dirac sea have the following 
commutational relations 
\[
\left[\r_1(-p);\r_1(p')\right]=\frac{pL}{2\pi}\delta_{p,p'}~~~~
\left[\r_2(p);\r_2(-p')\right]=\frac{pL}{2\pi}\delta_{p,p'},
\]
one can represent the Hamiltonian in the following form: 
\[
H=\frac{2\pi}{L}\sum_{p>0}\left(\r_1(p)\r_1(-p)+\r_2(-p)\r_2(p)\right)+ 
\l\sum_{p>0}\frac{2\pi}{L}\left(\r_1(p)\r_2(-p)+\r_1(-p)\r_2(p)\right).
\]
To evaluate the correlators in the system of finite length and make the 
connection with the conformal field theory predictions, one can proceed as follows. 
First one can define the lattice fields $n_{1,2}(x)$ with the help of the Fourier 
transform as 
\[
\rho_{1,2}(p)=\int_{0}^{L} dx e^{ipx}n_{1,2}(x),~~~~
n_{1,2}(x)=\frac{1}{L}\sum_p e^{-ipx}\rho_{1,2}(p)
\]
This fields have a physical meaning of the local number of the fermions 
above the Fermi level at the right and the left Fermi points. In terms of this 
fields the Hamiltonian has the following form:
\[
H=2\pi\sum_x\left(\frac{1}{2}(n_1^2(x)+n_2^2(x))+\lambda n_1(x)n_2(x)\right)
\]
Considering  the average distribution of the number of extra particles we obtain
the ground state energy in the form
\[
\Delta E=\frac{2\pi}{L}\left(\frac{1}{2}\left((\D N_1)^2+(\D N_2)^2\right)
+\l\D N_1 \D N_2 \right)
\]
where $\D N_{1,2}$ - are the numbers of additional particles at the two Fermi points. 
One can also rewrite the  ground state energy in the sector with the total number of
particles and the momentum $\D N=\D N_1+\D N_2,~~\D Q=\D N_1-\D N_2$
in such a way that the total Hamiltonian takes the following form (this form was first 
proposed in ref.\cite{H}): 
\be
H=u(\l)\sum_p|p|b_p^{+}b_p+\frac{\pi}{2L}u(\l)\left[\xi(\D N)^2+(1/\xi)(\D Q)^2\right],
\label{finite}
\ee
where the parameters $u(\l)=(1-\l^2)^{1/2}$ and $\xi=((1+\l)/(1-\l))^{1/2}$. 
Next one establishes the commutational relations for the fields $n_{1,2}(x)$: 
\[
\left[ n_{1}(x); n_{1}(y)\right]=-\frac{i}{2\pi}\delta'(x-y)~~~~
\left[ n_{2}(x); n_{2}(y)\right]=\frac{i}{2\pi}\delta'(x-y) 
\]
Then introducing the new variables $\n_{1,2}(x)=\sqrt{2\pi}~ n_{1,2}(x)$,
we have the following density of the Hamiltonian 
\be
H=\frac{1}{2}\left(\n_1(x)\n_1(x)+\n_2(x)\n_2(x)\right)
+ \lambda \n_1(x)\n_2(x).
\label{v}
\ee
We also have the following commutational relations
$\left[\n_1(x);\n_1(y)\right]=-i\delta^{\prime}(x-y)$.   
We then have the following conjugated field and the momenta:
\[
\pi(x)= -\frac{1}{\sqrt{2}}(\n_1(x)-\n_2(x));~~~~  
\d_x\phi(x)= \frac{1}{\sqrt{2}}(\n_1(x)+\n_2(x))
\]
In terms of these variables the Hamiltonian takes the following form: 
\be
H=\frac{1}{2}u(\l)\left[ (1/\xi)\pi^2(x)+ \xi(\d\phi(x))^2\right]
 = \frac{1}{2}u(\l)\left[ \hat{\pi}^2(x)+ (\d\hat{\phi}(x))^2\right], 
\label{h}
\ee
where 
\be
\pi(x)=\sqrt{\xi}~\hat{\pi}(x),~~~ 
\phi(x)=(1/\sqrt{\xi})\hat{\phi}(x).  
\label{canon}
\ee
The last equation (\ref{canon}) is nothing else as the canonical transformation,  
which is equivalent to the Bogoliubov transformation for the original operators 
$\rho_{1,2}(p)$. Next to establish the expressions for Fermions one should use 
the commutational relations $\left[a^{+}(x);\rho_{1}(p)\right]=-e^{ipx}a^{+}(x)$
and the same for $c^{+}(x)$. Note that these last relations were obtained using the 
expression with original fermions: 
$\rho_{1}(p)=\int dy e^{ipy}a^{+}(y)a(y)$. 
In this way we obtain the following expressions for fermionic operators:
\be
a^{+}(x)=K_1^{+}\frac{1}{\sqrt{2\pi\a}}
\exp\left(\frac{2\pi}{L}\sum_{p\neq0}\frac{\rho_1(p)}{p}e^{-ipx}e^{-\a|p|/2}\right)
=K_1^{+}\frac{1}{\sqrt{2\pi\a}}\exp\left(-i2\pi \tilde{N}_1(x)\right),
\label{fermions}
\ee
\[
c^{+}(x)=K_2^{+}\frac{1}{\sqrt{2\pi\a}}
\exp\left(-\frac{2\pi}{L}\sum_{p\neq0}\frac{\rho_2(p)}{p}e^{-ipx}e^{-\a|p|/2}\right)
=K_2^{+}\frac{1}{\sqrt{2\pi\a}}\exp\left(i2\pi \tilde{N}_2(x)\right),
\]
where the fields $\tilde{N}_{1,2}(x)$ are the analogs of the particle numbers 
with positions left to the point $x$  
and $K_{1,2}^{+}$ are 
the Klein factors - the operators 
which commute with the operators $\rho_{1,2}(p)$  
and create the single particle at the right (left) Fermi -points when 
acting on the ground state. The parameter $\a\to 0$ is introduced to 
perform the ultraviolet cutoff. The operators (\ref{fermions}) have the correct 
anticommutational relations.

It is assumed that there is an equality between the matrix elements 
of the specific 1D model for the low-energy states and the matrix elements 
of the corresponding operators over the corresponding eigenstates 
in the Luttinger liquid theory. 
The mapping between the two models is characterized by a single parameter 
$\xi$ which should be the same for both models in a sense of the 
equation (\ref{finite}) 
i.e. the constant 
$\xi$ in the effective Luttinger model is taken from the expression (\ref{finite}) 
for the original model. 
Thus the critical exponents in the asymptotics 
of the correlators for the 1D models are expressed in terms of the single 
parameter $\xi$ i.e. exhibit the universal behaviour. 
The constant $\xi$ determines the asymptotic behaviour 
of the correlators in the original model according to the prescriptions for the 
Luttinger model. 
This hypothesis is confirmed by the fact that the expression 
(\ref{finite}) is valid for the original model with the corresponding true 
speed of sound $v=u(\l)$. 
This was proved both for the continuous Bose or Fermi liquids and for 
the XXZ- spin chain \cite{H}.

We propose that the correspondence between the two models 
is not limited only by the correspondence between the critical exponents. 
There is an exact mapping between the operators and the eigenstates 
for the original and the Luttinger models. 
The matrix elements and the correlators in the original model are 
{\it equal} to the matrix elements and the correlators in the 
Luttinger model provided the operators are replaced by the corresponding 
operators in the Luttinger model and the eigenstates are replaced by the 
corresponding eigenstates. This correspondence for the low-lying states is 
valid if at the large distances the correlators are effectively described 
by the small energies and the contributions of the large energies are 
parametrized by the non-universal constants entering the definition of 
the operators in the Luttinger model. 
Thus for example the correlators at large distances can be calculated 
exactly in the framework of the effective low-energy Luttinger model. 
One can call these ideas by the extended bosonization concept, 
which means the exact mapping between the original and the Luttinger 
liquid models.

\vspace{0.2in}

{\bf 3. Lowest formfactors.} 

\vspace{0.2in}

Let us start with the derivation of the scaling relations for the lowest 
formfactors for the XXZ- quantum spin chain: 
\[
H=\sum_{i=1}^{L}\left(\s^{x}_i\s^{x}_{i+1}+\s^{y}_i\s^{y}_{i+1}
+\D\s^{z}_i\s^{z}_{i+1}\right),  
\]
where the sites $L+1$ and $1$ are coincide. First, let us establish the relation 
for the formfactors of $\s_i^{\pm}$- operators. The relations for the other operators 
($\s_i^{z}$) as well as for the formfactors for the other 1D systems can be 
obtained in a similar way. 
Calculation of finite - size corrections to the energy of the ground state for 
the XXZ- spin chain (see for example \cite{Kar}) leads to the expression  
(\ref{finite}) and allows one to obtain the parameter $\xi$ which leads to 
the predictions of critical indices according to the conformal field theory. 
The calculation gives the value $\xi=2(\pi-\eta)/\pi$, where the parameter $\eta$ 
is connected with the anisotropy parameter of the XXZ - chain as 
$\Delta=\cos(\eta)$. 
Using the Jordan-Wigner transformation $\sigma^{+}_x=a^{+}_x\exp(i\pi N(x))$,
where $a^{+}_x$ stands for the ``original'' lattice fermionic operator, 
$N(x)=\sum_{l=1}^{x-1}n_l$ ($n_l=a_l^{+}a_l$),   
and performing the obvious substitutions $N(x)\rightarrow x/2+N_1(x)+N_2(x)$ and
$a^{+}_x\rightarrow e^{-ip_Fx}a^{+}(x)+e^{ip_Fx}c^{+}(x),~~p_F=\pi/2$,  
we obtain after the canonical transformation (\ref{canon})
the expression for the leading term in the asympotics of the correlator for the XXZ -chain:
\be
\la\sigma^{+}_x\sigma^{-}_0\ra=\la a_x^{+}e^{i\pi N(x)}a_0\ra
\sim(-1)^x
\la0|e^{-i\pi\sqrt{\xi}(\tilde{N}_1(x)-\tilde{N}_2(x))}
e^{i\pi\sqrt{\xi}(\tilde{N}_1(0)-\tilde{N}_2(0))} 
|0\ra,
\label{leading}
\ee
where $\tilde{N}_{1,2}(x)$ - are corresponds to the free fields 
$\hat{\pi}(x)$, $\hat{\phi}(x)$, obtained after the transformation (\ref{canon}). 
To these operators correspond the new operators $\rho_{1,2}(p)$ and the new 
fermionic operators (quasiparticles). 
Averaging the product of exponents in bosonic operators 
for the expression (\ref{leading}) and using the properties of $\rho_{1,2}(p)$, 
$\la\rho_1(-p)\rho_1(p)\ra=\frac{pL}{2\pi}\theta(p)$ and 
$\la\rho_2(p)\rho_2(-p)\ra=\frac{pL}{2\pi}\theta(p)$, 
we get for the correlation function 
$G(x)=\la0|\sigma^{+}_{i+x}\sigma^{-}_{i}|0\ra$ the following sum in 
the exponent: 
\[
C~\exp\left(~\frac{\xi}{4}~\sum_{n=1}^{\infty}\frac{1}{n}e^{in(2\pi x/L)}
+ h.c.\right), 
\]
where $C$ - is some constant. Then using the formula   
$\sum_{n=1}^{\infty}\frac{1}{n}z^n=-\ln(1-z)$ and substituting the value
$\xi=2(\pi-\eta)/\pi$ we obtain the following 
expression for the XXZ - chain: 
\be
G(x)=C_0\frac{(-1)^x}{\left(L\sin(\frac{\pi x}{L})\right)^{\alpha}},
~~~\alpha=\frac{\xi}{2}=\frac{\pi-\eta}{\pi} ~~~(x>>1).
\label{coas}
\ee
Thus, although bosonization, which deals with the low-energy effective theory, 
is not able to predict the constant before the asymptotics, the critical exponent 
and the functional form are predicted in accordance with conformal field theory.

From the relation (\ref{leading}) it follows the following equation
(the local operators of the XXZ- chain should be represented as the 
local operators of the Luttinger model): 
\be
\s^{-}_{0}=C^{\prime}K_{1}e^{i\pi\sqrt{\xi}(\Nt_1(0)-\Nt_2(0))}+\ldots.  
\label{operator}
\ee
where $C^{\prime}$ is some constant and the dots stand for the different 
(subleading) operators. Eq.(\ref{operator}) should be understood in a 
sense of the correspondence between the XXZ and the Luttinger liquid 
models: the formfactors for the corresponding states should be equal 
to each other. In particular for the ground states of $M$ ($|t\ra$) and 
$M-1$ ($|\lambda\ra$) particles (up-spins) for the XXZ spin chain 
we obtain: 
\be
\la\lambda|\s^{-}_{0}|t\ra=
C^{\prime}\la0|e^{i\pi\sqrt{\xi}(\Nt_1(0)-\Nt_2(0))}|0\ra.  
\label{lowest}
\ee
This equation is valid if we assume that the eigenstate $\la\lambda|$ 
corresponds to the state $\la-1|_{(1)}$ i.e. the state with the 
absence of particle at the first branch of the Luttinger model. 
Calculating the average at the right-hand side of the equation 
(\ref{lowest}) we obtain $C^{\prime}=C(L/2\pi\a)^{\xi/4}$, where 
$C=\la\lambda|\s^{-}_{0}|t\ra$ is the value of the lowest formfactor. 
Next calculating the correlator $G(x)$ in the framework of the 
Luttinger model using the equation (\ref{operator}) we obtain the 
following result: 
\be
G(x)=(C^{\prime})^{2}\la0|e^{-i\pi\sqrt{\xi}(\tilde{N}_1(x)-\tilde{N}_2(x))}
e^{i\pi\sqrt{\xi}(\tilde{N}_1(0)-\tilde{N}_2(0))}|0\ra= 
\frac{C^2}{(2\sin(\pi x/L)^{\xi/2}}. 
\label{corr}
\ee
Note that the dependence of $G(x)$ on the parameter $\a$ is cancelled 
due to the dependence of the constant $C^{\prime}=C(L/2\pi\a)^{\xi/4}$ 
on $\a$ which indicates the independence of the scaling relations on 
the details of physics at high momenta which is model-dependent.  
Comparing the equation (\ref{corr}) with the equation (\ref{coas}) we 
obtain the desired relation between the lowest formfactor $C$ and 
the prefactor $C_0$: 
\be
C^2=\left(\frac{2}{L}\right)^{\xi/2}C_0.
\label{scaling}
\ee
For the case of the XX- spin chain ($\xi=1$) the relation between the 
formfactor and the prefactor (\ref{scaling}) coincides with the relation 
obtained in Ref.\cite{O} using the completely different method. 
Thus we see that while the prefactors are not the universal quantities 
the relations between them and the corresponding formfactors are universal.

Let us consider the density-density correlator for the XXZ- spin chain, 
namely $\Pi(x)=\la\s^{z}_{x}\s^{z}_{0}\ra$. 
Substituting the expressions for the original lattice Jordan-Wigner 
fermions into the density operator $n_x=a^{+}_{x}a_x$, we obtain the 
general expression for the density in terms of the Luttinger liquid 
operators: 
\be
n_x=\frac{1}{\sqrt{\pi}}\frac{1}{\sqrt{\xi}}\d_{x}\hat{\phi}(x)+ 
C^{\prime}_{1}e^{-i2p_{F}x}K_1^{+}K_{2}e^{-i2\pi(1/\sqrt{\xi})
\left(\Nt_1(x)+\Nt_2(x)\right)}+h.c.+\ldots,  
\label{oper}
\ee
where $p_F=\pi/2$, the operators $\Nt_{1,2}(x)$ correspond to the new 
fields $\hat{\pi}(x),\hat{\phi}(x)$ and the dots stand for the operators,
corresponding to the higher order terms in the expansion of the correlator. 
Consider the matrix element $\la t^{\prime}|n_0|t\ra=C_1$ , where $|t\ra$ -
is the ground state of the XXZ- spin chain and $|t^{\prime}\ra$ is the 
eigenstate which is obtained from the ground state by adding one particle 
at the right Fermi- point and removing one particle from the left Fermi- 
point. Taking the corresponding matrix element for both sides of the equation 
(\ref{oper}) we obtain: 
\be
\la t^{\prime}|n_0|t\ra=C_1=C_1^{\prime}
\la0|e^{-i2\pi(1/\sqrt{\xi})\left(\Nt_1(0)+\Nt_2(0)\right)}|0\ra.  
\label{me}
\ee
Calculating the average at the right- hand side side of Eq.(\ref{me}) 
we obtain the constant $C_1^{\prime}=C_1(L/2\pi\a)^{1/\xi}$. 
Calculating the correlator $\Pi(x)$ in the framework of the Luttinger 
liquid theory using the equation (\ref{oper}), we obtain the expression 
for the first two terms in the expansion in the form: 
\be 
\Pi(x)=-\frac{1}{2\xi(L\sin(\pi x/L))^2}+ e^{i2p_{F}x}
\frac{C_1^2}{\left(2\sin(\pi x/L)\right)^{2/\xi}}+h.c.. 
\label{dens}
\ee
Again the parameter $\a$ drops out of this equation. 
Comparing this equation with the general expression for the correlator: 
\be 
\Pi(x)=-\frac{1}{2\xi(L\sin(\pi x/L))^2}+ 
\frac{C_{10}\cos(2p_{F}x)}{\left(L\sin(\pi x/L)\right)^{2/\xi}}+\ldots,  
\label{density}
\ee
we obtain the scaling relation: 
\be
C_1^2=\frac{1}{2}\left(\frac{2}{L}\right)^{2/\xi}C_{10}. 
\label{scaldens}
\ee
The equation (\ref{scaldens}) was proved explicitly by direct 
computations for the XXZ spin chain in Ref.\cite{S1}. 

The examples presented above allow one to conclude the following. 
First, it is clear that in the same way the similar relations 
can be obtained for the other models of the 1D quantum liquids,  
in particular for the correlators of the continuous Bose- or 
Fermi- liquids. Second, the scaling relations can be easily 
generalized to the case of the lowest formfactors corresponding 
to the higher terms in the asymptotics of the correlators. 
The corresponding states are the states obtained from the ground 
state by moving an arbitrary number ($m$) of particles from the 
left to the right Fermi- points, which corresponds to the operators 
containing an additional powers of the operator $(a^{+}(x)c(x))$: 
$(a^{+}c)^m$. 
It is clear that in all the cases the results have the same form 
as the equations (\ref{scaling}), (\ref{scaldens}) with the 
corresponding critical exponent $\a(m)$. 
An explicit expressions for the correlators and the corresponding 
scaling relations for the 1D continuous Bose- and Fermi- liquids 
are presented in Ref.\cite{G}. 
For completeness we present these results in the Appendix A.  
Here we would like to stress once more 
that these results of Ref.\cite{G} can be equally well obtained 
in the framework of the bosonization approach.

\vspace{0.2in}

{\bf 4. Particle-hole formfactors.} 

\vspace{0.2in}

Let us consider the calculation of the low-energy particle-hole 
formfactors in the framework of the bosonization approach. 
As an example consider the formfactor of the operator $\s_0^{-}$ 
for the XXZ spin chain. Suppose we have the eigenstate 
$\la\lambda(p_i,q_i)|$ obtained from the ground state by creating 
the holes with the momenta $q_i$ and the particles with the 
momenta $p_i$ ($i=1,\ldots n$) located in the vicinity of the 
right Fermi- point.
The formfactor corresponding to 
this state has the following representation in terms of Luttinger 
liquid matrix element: 
\be
\la\lambda(p_i,q_i)|\s_0^{-}|t\ra=
C^{\prime}\la{p_i,q_i}|e^{i\pi\sqrt{\xi}(\Nt_1(0)-\Nt_2(0))}|0\ra, 
\label{ph}
\ee
where at the right-hand side 
$p_i>0$ and $q_i\leq 0$ are the positions of the particles 
and the holes at the first branch of the Luttinger liquid 
(i.e. correspond to the operators $a^{+}(x)$, $a(x)$) and the 
constant $C^{\prime}$ was introduced in Eq.(\ref{operator}). 
Calculating the average at the right-hand side of the 
equation (\ref{ph}) we obtain the following result: 
\be 
\la\lambda(p_i,q_i)|\s_0^{-}|t\ra= 
C\la{p_i,q_i}|e^{a\frac{2\pi}{L}\sum_{p>0}\frac{\rho_1(p)}{p}}|0\ra, 
~~~~a=-\sqrt{\xi}/2,
\label{Lph}
\ee
where $C$ is the value of the lowest formfactor 
$C=\la\l|\s_0^{-}|t\ra$. 
The average at the right- hand side of the equation (\ref{Lph}) 
was calculated in Ref.\cite{A} (see Appendix B):  
\be 
\la{p_i,q_i}|e^{a\frac{2\pi}{L}\sum_{p>0}\frac{\rho_1(p)}{p}}|0\ra
=F_{a}(p_i,q_i)=\det_{ij}\left(\frac{1}{p_i-q_j}\right)
\prod_{i=1}^{n}f^{+}(p_i)\prod_{i=1}^{n}f^{-}(q_i),
\label{F1}
\ee
where
\[
f^{+}(p)=\frac{\Gamma(p+a)}{\Gamma(p)\Gamma(a)}, ~~~~
f^{-}(q)=\frac{\Gamma(1-q-a)}{\Gamma(1-q)\Gamma(1-a)}. 
\]
In the equation (\ref{F1}) $p_i$ and $q_i$ are assumed to be 
integers (corresponding to the momenta $2\pi p_i/L$ and 
$2\pi q_i/L$) and $p_i>0$, $q_i\leq 0$.

Thus we have calculated the particle-hole formfactor in the form 
$\la\lambda(p_i,q_i)|\s_0^{-}|t\ra=CF(p_i,q_i)$, and have shown 
that the constant $C$ is in fact the lowest formfactor, which is 
not clear within the approach of Ref.\cite{G}. 
The same analysis can be performed for the formfactor 
$\la t^{\prime}(p_i,q_i)|n_0|t\ra=C_{1}F_{a}(p_i,q_i)$, where the 
constant $a$ in Eq.(\ref{F1}) is now equals $a=1/\sqrt{\xi}$, 
and the formfactors for the continuous Bose- and Fermi- liquids. 
Note that the dependence on the particles and holes momenta 
(\ref{F1}) can be easily obtained from the expressions for the 
formfactors in the case of the XX- spin chain in the form of 
the Cauchy determinant in Ref.\cite{O} ($\xi=1$, $a=-1/2$). 
We considered the formfactors corresponding to the particles 
and the holes located near the right Fermi- point. 
The same analysis can be performed for the particles 
and the holes located near the left Fermi- point (see below).  
Clearly, the total formfactor is a product of these two terms. 
Note that the corresponding particle-hole formfactors for the 
continuous Bose- liquid are given by the same formulas with 
the same value of the parameter $a$.

To sum up the formfactor series for the correlators in the 
framework of the approach of Ref.\cite{S2} (where the dependence 
on the particle and hole positions was found for the formfactors 
of the XXZ- spin chain) it is useful to have the formula for the 
sum: 
\be
\sum_{n}\sum_{p_i>0,q_i\leq0}|F_{a}(p_i,q_i)|^2 
e^{i(p-q)2\pi x/L}=\frac{1}{(1-e^{i2\pi x/L})^{a^2}}, 
\label{sum}
\ee
where $p=\sum_{i=1}^{n}p_i$, $q=\sum_{i=1}^{n}q_i$ and 
$p_i$, $q_i$ are integers. 
The result (\ref{sum}) can be easily obtained with the help of 
the following correlator: 
\[
G_{a}(x)=\la0|
e^{-a\frac{2\pi}{L}\sum_{p<0}\frac{\rho_1(p)}{p}e^{-ipx}} 
e^{a\frac{2\pi}{L}\sum_{p>0}\frac{\rho_1(p)}{p}}|0\ra= 
\frac{1}{(1-e^{i2\pi x/L})^{a^2}}, 
\]
which can be calculated either with the help of inserting 
of the complete set of intermediate states or by means of the 
standard formulas used in the bosonization approach. 
Thus the two sides of the equation (\ref{sum}) should be equal 
to each other. 

Finally, it would be interesting to verify the sum rule for 
the function $F(p_i,q_i)$ which can be obtained from the 
equation (\ref{sum}) by means of expanding it into the 
Fourier series: 
\[
\sum_{n}\sum_{p_i,q_i,p-q=m}|F_{a}(p_i,q_i)|^2= 
\frac{\Gamma(a^2+m)}{\Gamma(m+1)\Gamma(a^2)}, 
\]
where $p=\sum_{i=1}^{n}p_i$, $q=\sum_{i=1}^{n}q_i$.

Now let us consider the particle-hole excitations in the vicinity of the 
left Fermi-point. For the case of the operator $\s_0^{-}$ for the XXZ- 
spin chain the formfactor is given by the equation 
\be
\la\lambda(p_i,q_i)|\s_0^{-}|t\ra=
C^{\prime}\la{p_i,q_i}|e^{i\pi\sqrt{\xi}(\Nt_1(0)-\Nt_2(0))}|0\ra, 
\label{ph2}
\ee
where now the momenta $p_i<0$, $q_i\geq0$ are the positions of the particles and 
the holes at the second branch of the Luttinger liquid, i.e. the momenta with 
respect to the left Fermi-point. 
Calculating the average at the right-hand side of Eq.(\ref{ph2}) we obtain: 
\be 
\la\lambda(p_i,q_i)|\s_0^{-}|t\ra= 
C\la{p_i,q_i}|e^{c\frac{2\pi}{L}\sum_{p<0}\frac{\rho_2(p)}{p}}|0\ra, 
~~~~c=\sqrt{\xi}/2,
\label{Lph2}
\ee  
where $C$ is again the value of the lowest formfactor $C=\la\l|\s_0^{-}|t\ra$. 
The average at the right-hand side of Eq.(\ref{Lph2}) is calculated in the 
Appendix B: 
\be 
\la{p_i,q_i}|e^{c\frac{2\pi}{L}\sum_{p<0}\frac{\rho_2(p)}{p}}|0\ra
=F_{c}(p_i,q_i)=\det_{ij}\left(\frac{1}{p_i-q_j}\right)
\prod_{i=1}^{n}f^{+}(p_i)\prod_{i=1}^{n}f^{-}(q_i),
\label{F2}
\ee
where now 
\[
f^{+}(p)=\frac{\Gamma(-p-c)}{\Gamma(-p)\Gamma(1-c)},~~~~~~
f^{-}(q)=\frac{\Gamma(1+q+c)}{\Gamma(1+q)\Gamma(c)}. 
\]
In the equation (\ref{F2}) $p_i$ and $q_i$ are assumed to be 
integers (corresponding to the momenta $2\pi p_i/L$ and 
$2\pi q_i/L$) and $p_i<0$, $q_i\geq 0$.
The same analysis can be performed for the formfactor 
$\la t^{\prime}(p_i,q_i)|n_0|t\ra=C_{1}F_{c}(p_i,q_i)$, 
where the particles and the holes are located near the left Fermi-point  
and the constant $c$ in Eq.(\ref{F2}) is now equals $c=-1/\sqrt{\xi}$, 
and the formfactors for the continuous Bose- and Fermi- liquids.
For example for the continuous Bose- liquid we obtain the same expression 
as for the XXZ- spin chain: 
\[
\la\l(p_i,q_i)_{1(2)}|\phi(0)|t\ra=CF_{a(c)}(p_i,q_i), 
\]
with $a=-\sqrt{\xi}/2$, $c=\sqrt{\xi}/2$, where the field $\phi(x)$ 
corresponds to the Bose- particles and $C$ is the value of the lowest 
formfactor $\la\l|\phi(0)|t\ra$. 
The formfactor $\la t^{\prime}(p_i,q_i)|\rho(0)|t\ra$ of the 
density operator $\rho(x)$ for the Bose- and Fermi- liquids also 
coincides with the corresponding particle-hole formfactor 
$\la t^{\prime}(p_i,q_i)|n_0|t\ra$ for the XXZ- spin chain.

To sum up the formfactor series for the correlators it is useful 
to have the formula for the sum: 
\be
\sum_{n}\sum_{p_i<0,q_i\geq0}|F_{c}(p_i,q_i)|^2 e^{i(p-q)2\pi x/L}=
\frac{1}{(1-e^{-i2\pi x/L})^{c^2}}, 
\label{sum2}
\ee
where $p=\sum_{i=1}^{n}p_i$, $q=\sum_{i=1}^{n}q_i$ and 
$p_i$, $q_i$ are integers. 
The result (\ref{sum2}) can be easily obtained with the help of 
the following correlator: 
\[
G_{c}(x)=\la0|
e^{-c\frac{2\pi}{L}\sum_{p>0}\frac{\rho_2(p)}{p}e^{-ipx}} 
e^{c\frac{2\pi}{L}\sum_{p<0}\frac{\rho_2(p)}{p}}|0\ra= 
\frac{1}{(1-e^{-i2\pi x/L})^{c^2}}, 
\]
which can be calculated again either with the help of inserting 
of the complete set of intermediate states or by means of the 
standard formulas used in the bosonization approach. 
Thus the two sides of the equation (\ref{sum2}) should be equal 
to each other. 
Let us mention that if one have the formulas (\ref{sum}), 
(\ref{sum2}) and the form of the particle-hole formfactors then 
one can prove the scaling relations for the lowest formfactors. 
Alternatively, if one have the formulas (\ref{sum}), (\ref{sum2}) 
and the scaling relations for the lowest formfactors, 
then assuming the formulas (\ref{Lph}), (\ref{Lph2}) one can 
prove that the constant $C$ in this formulas is in fact the 
lowest formfactor.

Finally let us calculate the particle-hole formfactors for the continuous 
Fermi-liquid. We have the following expression for the Fermi-operator: 
\[
\psi(0)=C^{\prime}K_{1}e^{i\pi\sqrt{\xi}(\Nt_1(0)-\Nt_2(0))+ 
i\pi(1/\sqrt{\xi})(\Nt_1(0)+\Nt_2(0))}+ 
\]
\[
~~~~~~C^{\prime}K_{2}e^{i\pi\sqrt{\xi}(\Nt_1(0)-\Nt_2(0))- 
i\pi(1/\sqrt{\xi})(\Nt_1(0)+\Nt_2(0))}+\ldots, 
\]
where the dots stand for the different powers of the Klein factors.   
The constant $C^{\prime}=C(L/2\pi\a)^{(1/4)(\xi+1/\xi)}$ is related 
to the lowest formfactors 
\[
C=\la\l_1|\psi(0)|t\ra=\la\l_2|\psi(0)|t\ra, 
\]
where $\la\l_1|$ ($\la\l_2|$) is the eigenstate corresponding to the 
absense of particle at the right(left) Fermi-point. 
We then have the following expressions for the particle-hole 
formfactors corresponding to the right(left) Fermi-point: 
\[
\la\l_{1}(p_i,q_i)_{1}|\psi(0)|t\ra=CF_{a}(p_i,q_i), ~~~~~
a=-\frac{1}{2}(\sqrt{\xi}+1/\sqrt{\xi}), 
\]
\[
\la\l_{1}(p_i,q_i)_{2}|\psi(0)|t\ra=CF_{c}(p_i,q_i), ~~~~~
c=\frac{1}{2}(\sqrt{\xi}-1/\sqrt{\xi}), 
\]
\[
\la\l_{2}(p_i,q_i)_{1}|\psi(0)|t\ra=CF_{a}(p_i,q_i), ~~~~~
a=-\frac{1}{2}(\sqrt{\xi}-1/\sqrt{\xi}), 
\]
\[
\la\l_{2}(p_i,q_i)_{2}|\psi(0)|t\ra=CF_{c}(p_i,q_i), ~~~~~
c=\frac{1}{2}(\sqrt{\xi}+1/\sqrt{\xi}).  
\] 
For the formfactors corresponding to the eigenstates with 
the particles and the holes located at both sides of the Fermi-interval 
we have the product of these two functions $F_{a}(p_i,q_i)$ and 
$F_{c}(p_i,q_i)$.

In conclusion, 
for various one-dimensional quantum liquids in the framework of the Luttinger model 
(bosonization) we established the relations between the prefactors of the 
correlators and the formfactors of the corresponding local operators. 
The physical reason for existing of these relations is that in 1D only 
the low-lying particle-hole excitations contribute to the power-law 
asymptotics of the correlators. 
In fact, it turns out that in the 1D models there is the separation of energies. 
The contributions of large energies are all included in the non-universal 
constants in front of the operators of the effective low-energy Luttinger model, 
while the contributions of small energies are described by the operators 
of the Luttinger liquid. 
Since the sums in the equations (\ref{sum}), (\ref{sum2}) are convergent,
at large distances the correlator is determined 
by the small energies so it can be calculated exactly in the framework of the 
effective Luttinger liquid theory. 
The derivation of the scaling relations in the framework 
of the bosonization procedure allows one to substantiate the prediction for the 
formfactors corresponding to the low-lying particle-hole excitations. 
Let us stress once more that the relations of the type (\ref{Lph}) can be 
obtained only in the framework of the bosonization procedure, since only within 
this approach one can see that the constant $C$ in Eq.(\ref{Lph}) is in fact 
the lowest formfactor. 
Second, using the so called extended bosonization concept introduced in the 
present paper (exact mapping of the original model to the Luttinger liquid 
model) we derived the expressions for the particle-hole formfactors. 
Let us note that these results could be obtained also from the correspondence 
of the two correlators of Eq.(\ref{leading}).
We obtained an explicit expressions for the particle-hole formfactors 
both for the XXZ- spin chain and the continuous Bose- and Fermi- liquids 
including the particle-hole formfactors with the excitations corresponding 
to the left Fermi- point.  
We also obtained   
the formulas for the summation over the particle-hole states corresponding to the 
power-law asymptotics of the correlators.

\vspace{0.3in}

{\bf Appendix A.} 

\vspace{0.2in} 

Here we present without derivation the scaling relations for the 
formfactors for the continuous Bose- and Fermi- liquids and the XXZ- 
spin chain including the relations for the higher order terms 
in the asymptotics of the correlators. The results for the 
Bose-liquid have exactly the same form as the results for the 
XXZ- spin chain. We denote by $\phi(x)$ ($\psi(x)$) the fields 
corresponding to the Bose (Fermi) particles and by $\rho(x)$ the 
density operator. The general expressions 
for the expansions of the Bose- field and the density operators are: 
\[
\phi(x)=\sum_{m}C_{m}^{\prime}e^{-i2p_{F}mx}
e^{i\pi\sqrt{\xi}(\Nt_1(x)-\Nt_2(x))} 
e^{-i2\pi m(1/\sqrt{\xi})(\Nt_1(x)+\Nt_2(x))}, 
\]
\[
\psi(x)=\sum_{m}B_{m}^{\prime}e^{i(2m+1)p_{F}x}
e^{i\pi\sqrt{\xi}(\Nt_1(x)-\Nt_2(x))} 
e^{i\pi(2m+1)(1/\sqrt{\xi})(\Nt_1(x)+\Nt_2(x))}, 
\]
\[
\rho(x)=\frac{1}{\sqrt{\pi}}\frac{1}{\sqrt{\xi}}\d_{x}\hat{\phi}(x)+ 
\sum_{m\neq0}A_{m}^{\prime}e^{-i2p_{F}mx} 
e^{-i2\pi m(1/\sqrt{\xi})(\Nt_1(x)+\Nt_2(x))},
\]
where $p_F=\pi\rho_0$ is the Fermi- momentum (here we omit the 
Klein factors, for the correlators one should take the averages 
of the terms, corresponding to the same harmonics (same $m$)). 
Exactly the same expressions hold for the operator $\s^{-}_x$ and 
$\s^{z}_x$ for the XXZ- spin chain ($p_F=\pi/2$) 
The general expressions for the equal-time correlators have the form: 
\[
G_B(x)=\la\phi^{+}(x)\phi(0)\ra= \sum_{m\geq0}C_{m}
\frac{\cos(2p_{F}mx)}{\left(L\sin(\pi x/L)\right)^{\xi/2+m^{2}(2/\xi)}}, 
\]
\[
G_F(x)=\la\psi^{+}(x)\psi(0)\ra= 
\sum_{m\geq0}B_{m}
\frac{\sin((2m+1)p_{F}x)}{\left(L\sin(\pi x/L)\right)^{\xi/2+(2m+1)^2/2\xi}}, 
\]
\[
\Pi(x)=\la\rho(x)\rho(0)\ra=\rho_0^2-\frac{1}{2\xi(L\sin(\pi x/L))^2}+
\sum_{m\geq 1}A_{m}
\frac{\cos(2p_{F}mx)}{\left(L\sin(\pi x/L)\right)^{(2/\xi)m^2}}. 
\]
The expressions for $G_B(x)$ and $\Pi(x)$ also hold for the XXZ- spin 
chain (the correlators $G(x)$ and $\Pi(x)$). Then the scaling 
relations for the lowest formfactors have the following form: 
\[
|\la\l(m)|\phi(0)|t\ra|^2=\frac{(-1)^{m}C_{m}}{2-\delta_{0,m}}
\left(\frac{2}{L}\right)^{\xi/2+m^{2}(2/\xi)}, 
\]
\[
|\la\l(m)|\psi(0)|t\ra|^2=\frac{(-1)^m B_{m}}{2}
\left(\frac{2}{L}\right)^{\xi/2+(2m+1)^2/2\xi}, 
\]
\[
|\la t(m)|\rho(0)|t\ra|^2=\frac{A_{m}}{2}
\left(\frac{2}{L}\right)^{(2/\xi)m^2}, 
\]
where $|\l(m)\ra$ is the eigenstate with the number of particles 
equal to $M-1$ and with $m$ particles removed from the right and 
created at the left Fermi- point, and $|t\ra$ is the ground state 
of $M$ particles ($\rho_0=M/L$). 
For fermions the formfactors $\la\l_{1,2}|\psi(0)|t\ra$ introduced 
above correspond to the formfactor $\la\l(0)|\psi(0)|t\ra$ from the 
last equations and the higher formfactors $\la\l(m)|\psi(0)|t\ra$ 
($m>0$) correspond to the states with the momentum $\pm(2m+1)p_F$. 
The two eigenstates with an opposite momentum correspond to the 
two different harmonics in the factor $\sin((2m+1)p_{F}x)$. 
For bosons the eigenstate $|\l(0)\ra$ is not degenerate. 
For the case of the XXZ- spin 
chain the particles correspond to the up-spins and we have the 
same formulas as the formulas for the bosons 
($\phi(0)\rightarrow\s_0^{-}$, $\rho(0)\rightarrow\s_0^{z}$). 
The factors $(-1)^m$ in the last equations appear in the process 
of the calculations of the corresponding averages for the 
correlators at $m\neq0$.

\vspace{0.2in}

{\bf Appendix B.} 

\vspace{0.2in} 

Let us calculate the matrix elements (\ref{F1}), (\ref{F2}) 
which correspond to the particle-hole formfactors. Let us begin 
with the matrix element (\ref{F1}) corresponding to the first branch 
of the Luttinger model \cite{A}. We start with the matrix element for 
the operators in the coordinate space $\la0|a^{+}(x)a(y)e^{B_a}|0\ra$, 
$B_a=a(2\pi/L)\sum_{p>0}\rho_1(p)/p$. Using the formula 
$Ae^{B}=e^{B}(\sum_{n}(1/n!)\left[A,B\right]_n)$, one can commute 
the exponent $e^{B_a}$ to the left. Thus we obtain the following equation: 
\[
\la0|a^{+}(x)a(y)e^{B_a}|0\ra= 
\left(\frac{1-e^{i\xp}}{1-e^{i\yp}}\right)^{a}\la0|a^{+}(x)a(y)|0\ra=
\frac{1}{L}\left(\frac{1-e^{i\xp}}{1-e^{i\yp}}\right)^{a}
\frac{e^{i\yp}}{e^{i\yp}-e^{i\xp}}, 
\]
where $\xp=2\pi x/L$, $\yp=2\pi y/L$. Then for the derivative we obtain 
\[
(i\partial_x+i\partial_y)\la0|a^{+}(x)a(y)e^{B_a}|0\ra= 
-a\frac{2\pi}{L^2}\left(1-e^{i\xp}\right)^{a-1}e^{i\yp}
\left(1-e^{i\yp}\right)^{-(a+1)}. 
\]
Calculating the Fourier transform for both sides of this equation we obtain 
for the matrix element $\la0|a^{+}_{q}a_{p}e^{B_a}|0\ra$ the expression 
(\ref{F1}) for $n=1$ with the factors 
\[
f^{+}(p)=a\int_0^{2\pi}\frac{dy}{2\pi}e^{-i(p-1)y}
\left(1-e^{iy}\right)^{-(a+1)}, ~~~~p>0, 
\]
\[
f^{-}(q)=\int_0^{2\pi}\frac{dx}{2\pi}e^{iqx} 
\left(1-e^{ix}\right)^{a-1}, ~~~~q\leq0,  
\]
where $p$ and $q$ are integers. Calculating the integrals and using the 
Wick's theorem we obtain the equation (\ref{F1}) with the functions 
$f^{+}(p)$, $f^{-}(q)$ presented in the text. 

The derivation of the equation (\ref{F2}) is similar. We consider the 
matrix element $\la0|c^{+}(x)c(y)e^{B_c}|0\ra$ with 
$B_c=c(2\pi/L)\sum_{p<0}\rho_2(p)/p$. Commuting $e^{B_c}$ to the left 
we obtain: 
\[
\la0|c^{+}(x)c(y)e^{B_c}|0\ra= 
\left(\frac{1-e^{-i\yp}}{1-e^{-i\xp}}\right)^{c}\la0|c^{+}(x)c(y)|0\ra=
\frac{1}{L}\left(\frac{1-e^{-i\yp}}{1-e^{-i\xp}}\right)^{c}
\frac{e^{-i\yp}}{e^{-i\yp}-e^{-i\xp}}. 
\] 
Calculation of the derivatives gives 
\[
(i\partial_x+i\partial_y)\la0|c^{+}(x)c(y)e^{B_c}|0\ra=
-c\frac{2\pi}{L^2}\left(1-e^{-i\yp}\right)^{c-1}e^{-i\yp}
\left(1-e^{-i\xp}\right)^{-(c+1)}.
\]
Calculating the Fourier transform we obtain for the matrix element 
$\la0|c^{+}_{q}c_{p}e^{B_c}|0\ra$ the expression (\ref{F2}) for 
$n=1$ with the factors
\[
f^{+}(p)=\int_0^{2\pi}\frac{dy}{2\pi}e^{-i(p+1)y}
\left(1-e^{-iy}\right)^{c-1}, ~~~~p<0, 
\]
\[
f^{-}(q)=c\int_0^{2\pi}\frac{dx}{2\pi}e^{iqx}
\left(1-e^{-ix}\right)^{-(c+1)}, ~~~~q\geq0. 
\]
Calculating the integrals we obtain the equation (\ref{F2}) with the 
functions $f^{+}(p)$, $f^{-}(q)$ presented in the text.

\end{document}